\documentclass{article}
\usepackage{mrs2005,epsfig}
\begin{document} 
\title{Some thoughts and some new results about the Butcher--Oemler 
effect} 
 
\author{S. Andreon}
\affil{INAF--Osservatorio Astronomico di Brera, Milano, Italy}

\begin{abstract}
These short notes collect some thoughts about the way
Butcher--Oemler-like and evolutionary studies are performed. We
emphasize the shortcomings of several overlooked technical ingredients in the
above type of studies, with the hope of making a useful cookbook
for future works. We also briefly report new results of an ongoing 
Butcher--Oemler study at intermediate redshift.
\end{abstract} 
 
\section{Introduction} 

One of the central issues in extragalactic astronomy is to understand
how and when galaxies evolve, and whether and how their evolution is
related to environment. There is no consensus about the above
topics, even considering one single standard observable,  such as the
fraction of blue galaxies: from one side there are several claims about the
existence of a Butcher--Oemler (BO) effect (e.g.  Butcher \& Oemler
1985, Rakos \& Shombert 1995),  on the other side, some studies do not
detect any Butcher-Oemler effect (e.g. Andreon, Lobo \& Iovino 2004,
Ellingson et al. 2001), and/or cast doubts about the previously found 
results (e.g. Andreon, Lobo \& Iovino 2004).

We suggest here some thoughts about some current works
addressing evolutionary studies, often in a BO-like style, mostly of
technical nature, but with a strong impact on the final result. Details can be
found in Andreon \& Ettori (1999), Andreon, Lobo \& Iovino (2004),
Andreon et al. (2005). Although some considerations presented here seems
obvious, they are sometime overlooked.

\section{Problems}

\subsection{Cluster--related selection effect}

It is well known that clusters are not an homogenous family of objects, all
characterized by a single value of the fraction of blue galaxies. For example,
the Virgo cluster is much richer in blue galaxies than Coma (e.g. Butcher \&
Oemler 1984). Therefore, the comparison of clusters at different redshifts, 
i.e. any claim about the existence of a BO effect, relies on the assumption
that what is compared is the {\it same} cluster at different ages. Andreon \&
Ettori (1999) clarified, instead, that because of the lack of appropriate
cluster samples, clusters of different masses at different redshifts are
usually compared (see Figure 1), much like ``comparing unripe apples with ripe
oranges in understanding how fruit ripens". If one want to study the evolution
of galaxy properties, then the cluster sample should be carefully chosen and
cluster selection effects should be attentively addressed because of the known
correlation between cluster and galaxies properties.

Allington-Smith et al. (1993) show that the optical selection of clusters is
prone to produce a biased - hence inadequate - sample for studies on evolution
since at larger redshifts it naturally favours the inclusion in the sample of
clusters with a significant fraction of blue galaxies. Andreon, Lobo \& Iovino
(2004)  show that high redshift  clusters with a large fraction of blue
galaxies are over-represented in optical cluster catalogs by a factor two, with
respect to identical clusters but without a bursting population. Therefore,
optically selected clusters should be use with extreme caution in BO-like
studies. 


Clusters have various size and different values of the fraction of blue
galaxies at different clustercentric radii (outskirts are richer in blue
galaxies). Therefore, a comparison of clusters having different sizes cannot 
use fractions of blue galaxies measured at the same physical size (say, 1 Mpc),
because such a radius reaches the outskirts of poor clusters (rich in blue
galaxies), but enclosed the central regions only of rich clusters (poor in blue
galaxies). This is one of the reason why Butcher \& Oemler
(1984) suggested the use of an adaptive aperture.

\begin{figure}  
\vspace*{1.25cm}  
\begin{center}
\epsfig{figure=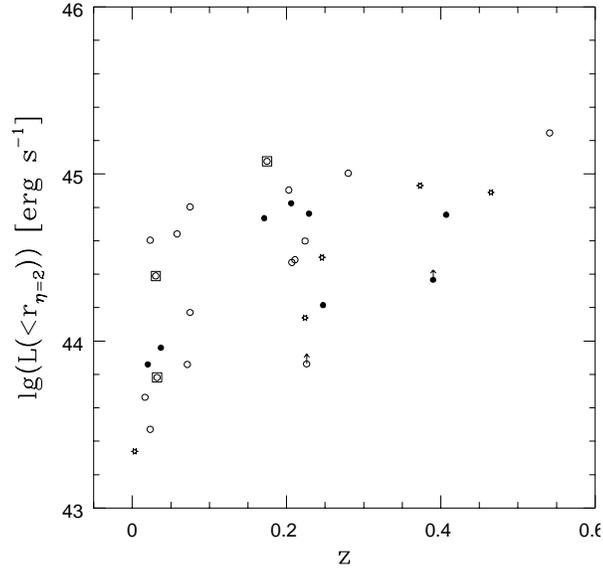,width=8cm,clip}
\end{center}
\vspace*{0.25cm}  
\caption{X-rax luminosity as a function of redshift for Butcher-Oemler
clusters. X-rax luminosity (and therefore mass) increases
with redshift because the way the cluster sample is selected: 
in fair samples luminosity stays constant or marginally decreases. 
Error bars have the size of
the data points. Figure taken from Andreon \& Ettori (1999).
} 
\end{figure}

%
%
\begin{figure}  
\vspace*{1.25cm}  
\begin{center}
\epsfig{figure=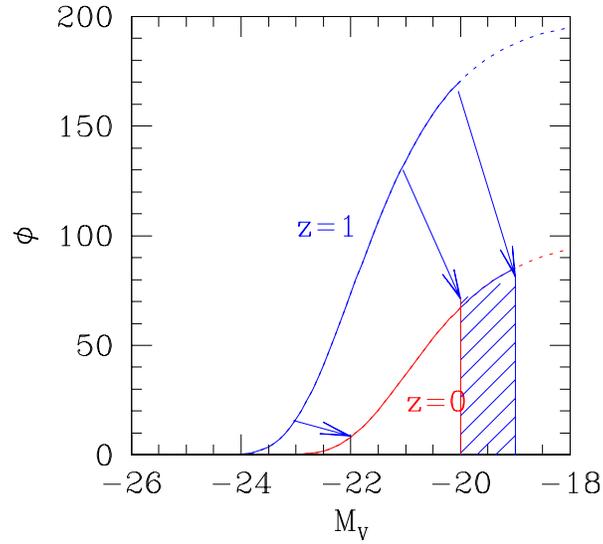,width=8cm,clip}
\end{center}
\vspace*{0.25cm}  
\caption{Luminosity function at $z=1$ (top curve)
and at $z=0$ (bottom curve) with arbitrary normalization. 
The shaded region is formed
by objects brighter than $M=-20$ mag at high
redshift (i.e. included in the studied sample), 
but faded below the $M=-20$ mag threshold at low redshift (i.e. excluded
in the low-redshift sample). } 
\end{figure} 

\subsection{Galaxy evolution: luminosity}

It is well known that galaxies evolve, i.e. that they do not take a
constant value of the absolute magnitude all along their life. Nevertheless,
in the selection of galaxies to be compared, most studies adopted a
fixed and unique absolute magnitude limit at all redshifts (the
Butcher--Oemler prescription is $M_V=-20$ mag). Between $z=1$ and $z=0$
the absolute luminosity of a typical galaxy changes by about one mag
(depending somewhat on filter and colour), which means that by selecting
a fixed and unique absolute magnitude limit at all redshifts we are not
selecting the same galaxy at two different look--back times: galaxies
in the faintest magnitude bin at high redshift are included in the
high-redshift sample, but their luminosity evolution will make them
fainter than the compared galaxies in the low redshift sample, as
sketched in Fig 2. Therefore, the choice of a fixed and unique absolute
magnitude limit induces a drift across the luminosity border. This drift
is  an import one:  for the above limiting magnitude, what drift out
from the sample is about 40 \% of what remains in the sample. 

The choice of a fixed and unique absolute  magnitude limit does not select
the same object at different redshift, and, because the know correlation
between luminosity, color, mass and size, may bring in
redshift--dependent biases on these quantities. These trends are
selection effects, however, not a measure of the evolution of galaxy
properties, and therefore are of nuisance, not of interest. We suggest
that if we want to focus on the evolution of galaxy properties one
should minimize the drift by choosing, for example, an evolving limiting
magnitude, as did  Ellingson et al. (2001), de Propris et al. (2003), 
Andreon, Lobo \& Iovino (2004),  Andreon et al. (2005) in their
studies. 

%
%
\begin{figure}  
\vspace*{1.25cm}  
\begin{center}
\epsfig{figure=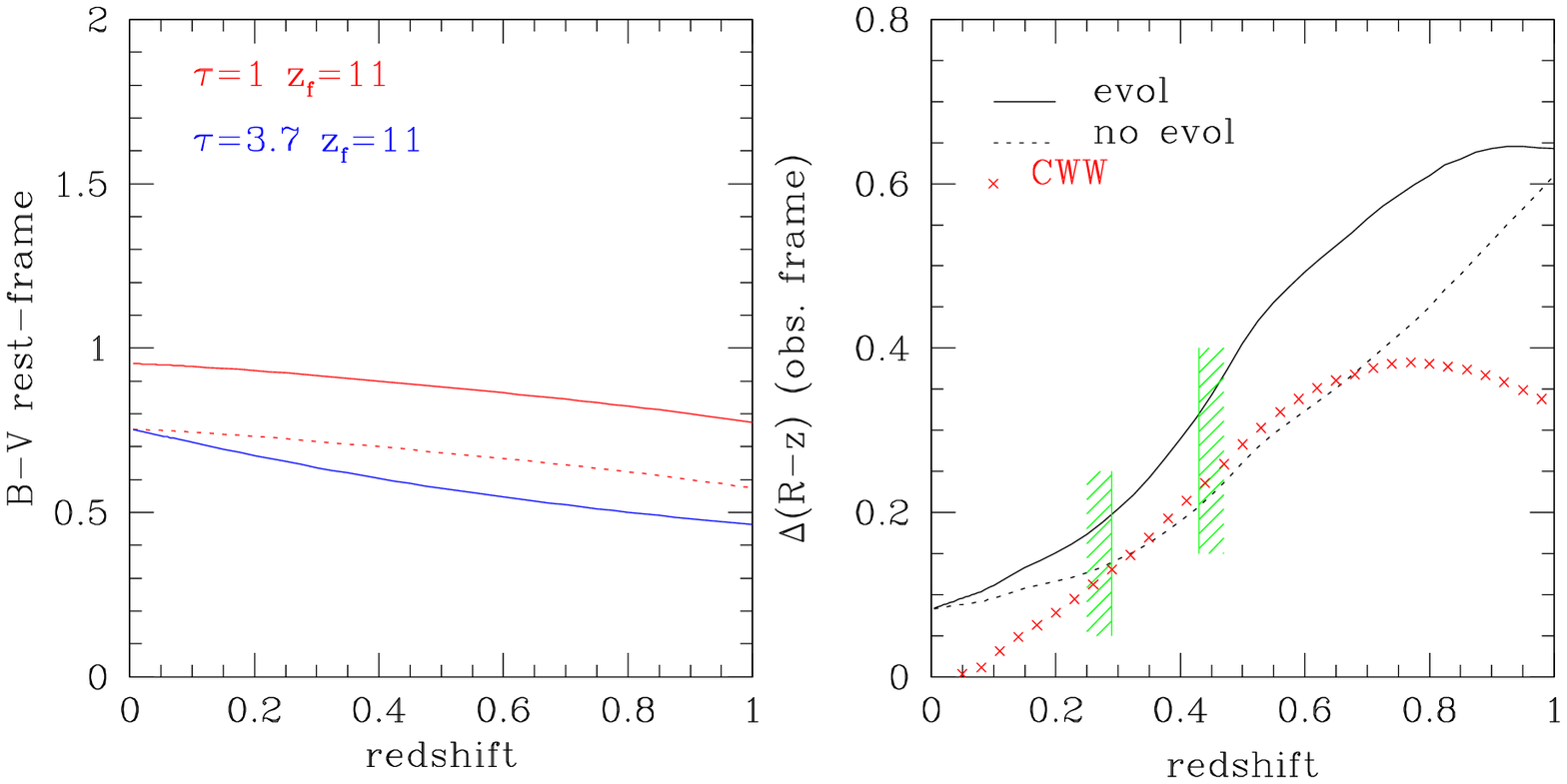,width=7cm,clip}  
\end{center}
\vspace*{0.25cm}  
\caption{  
Rest--frame $B-V$ colour of a $\tau=1$ Gyr and
$z_f=11$ stellar population (top red line, mimicking an E), and a 
stellar population having the same $z_f$ but being 0.2 mag bluer
today, i.e. $\tau=3.7$ Gyr, $z_f=11$  (bottom blue continuos line).
The dotted line show the Butcher--Oemler prescription. Objects below
the dotted line and above the continuous line drift from one
class to the other along their evolution. 
} 
\end{figure}

\subsection{Galaxy evolution: colour}
 
It is well known that the colour of the galaxies changes during their life.
Nevertheless, some studies adopted a fixed and unique absolute colour value to
define when a galaxy should be called ``blue" (e.g. Rakos \& Shombert 1995,
Faber et al. 2005, Nuijten et al. 2005). Let us consider, for example, as
definition of blue the following: $B-V<0.8$ mag (rest-frame). At $z\ga 1$ all
galaxies are blue, simply because the universe at $z\ga 1$ is not old enough to
allocate a redder population (unless an uncommon value of extinction or
metallicity is considered). Using the above definition of blue, the fraction of
blue galaxies does not tell about the relative evolution of red and blue
galaxies, simply because that fraction is obliged to be equal to 1 at $z>1$ and
to tend to 1 at $z\sim1$ {\it by definition}, i.e. independently on the
relative evolution of red and blue galaxies. We discourage the use of similar
definition of ``blue".

The Butcher \& Oemler (1985) definition of ``blue" is much better than the
above definition: galaxies are defined blue if they are at least 0.2 mag bluer
than early--type galaxies {\it at the same redshift}, i.e. $\Delta (B-V)=0.2$
mag. This definition can be improved: star aging modifies the  $\Delta (B-V)$ as
a function of the galaxy age (or look-back time): a galaxy having $\Delta
(B-V)=0.2$ mag {\it today} has not $\Delta (B-V)=0.2$ mag at a different
redshift, as shown in Figure 3. Objects below the dotted line and above the
continuous line drift from one class to the other as time goes.  Since the
choice of a fixed $\Delta$ allows a possible drift from one class to the other,
and assuming that a redshift dependence is found for the fraction of blue
galaxies, does the above  tell us something about the relative evolution of red
and blue  galaxies? It may merely reflect a selection bias related to the way
galaxies are divided in colour classes: a  class naturally gets contaminated by
the other one, and galaxies classes are mixed with different proportions at
different redshifts whit such a definition of ``blue".

We believe that adopting an evolving $\Delta (B-V)$ is preferable,
because it minimize the drift from one class to another. This solution
is model dependent, however. A better solution is to track some observed
feature of the colour distribution, such as the valley between red and
blue galaxies (e.g. Strateva et al. 2001) at all redshifts and use it to
divide galaxies in colour classes.

\subsection{The nightmare of the statistical analysis} 

Clusters of galaxies appear projected over foreground or background galaxies.
However, in most occasions, we ignore which galaxies are cluster members and
which ones are interlopers. Therefore, the background contribution is estimated
in some way (for example in a reference line of sight) and subtracted
statistically. In Andreon, Lobo \& Iovino (2004) we  introduce a first
discussion about the difficult task of measuring the error on the fraction of
blue galaxies   showing that at least some previous works have underestimated
errors and, by consequence, overstated the evidence for the BO effect. The task
has been further elaborated (and a rigorous solution has been found) in
d'Agostini (2004) (see also Andreon et al. 2005). We defer interested readers
to the above papers, remarking only the  complexity of the correct statistical
analysis. Let us focus in the apparent simple problem of the definition of the
fraction of blue galaxies. Intuitively (and according to Butcher \& Oemler
1985) the fraction of blue galaxies is the ratio between the number of blue
galaxies over the total number of galaxies. This is also the maximum likelihood
estimate (or best estimate) of the fraction of blue galaxies. However, the large
majority of papers studying the Butcher-Oemler effect find some clusters with
a negative fraction of blue galaxies  (e.g. Butcher \& Oemler 1984 and Wake et
al. 2005). As a physicians, negative values of the blue fraction  should
make us unhappy, because, for sure, no cluster has a negative number
of blue galaxies, thus suggesting that a better definition of fraction of blue
galaxies should exist (for example one that does not take impossible values).
As statisticians and mathematicians, negative values of the fraction of blue
galaxies cannot be best (maximum likelihood) estimates, because every value in
the 0 to 1 range is a better (more likely) estimate of the fraction of blue
galaxies than the quoted (impossible!) negative value. 

The point is that in presence of small number statistic,  an important
background or a physical border ($f_b \sim0$ or $\sim1$) it is dangerous to 
stick on intuition and on the
simple statistical rules (like the formula of summing errors in
quadrature,  or maximum likelihood estimates), because  the use
of simple rules may lead to un-robust results. A 
rigorous statistical analysis provides, instead,
results of known reliability.

\begin{figure}  
\vspace*{1.25cm}  
\begin{center}
\epsfig{figure=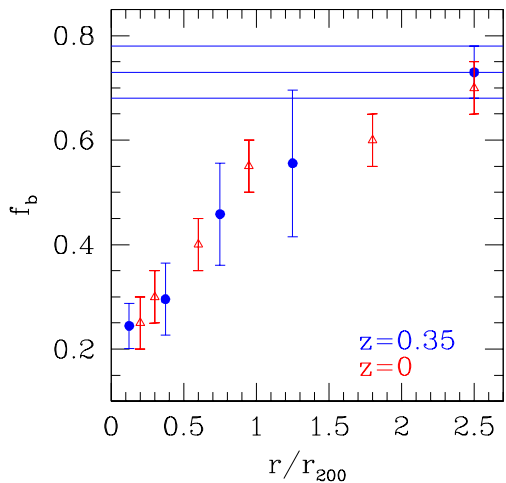,width=7cm,clip}  
\end{center}
\vspace*{0.25cm}  
\caption{Clustercentric dependence of the fraction of blue
galaxies. Filled dots represent the fraction of blue galaxies 
at $z\sim 0.35$. $z\sim0.35$ field value is 
arbitrarily set at $r/r_{200}=2.5$ for display purpose. 
Open triangles correspond to the fraction of galaxies with normalized
star formation rates larger than 1 solar mass per year
from Lewis et al. (2002) at $z\sim0$. Figure taken from Andreon et al.
(2005)} 
\end{figure}

\section{Results} 

The above difficulties are no unsurmountable problems.  Appropriate cluster
samples are being built, for example from x-ray surveys: x-ray luminosity is a
tracer of cluster mass, and such a selection should allow to select similar
clusters at different redshifts. The problems raised by the difficulty of the
statistical analysis are now surmounted. The remaining issues have already been
addressed in published papers although not every paper account for every issue.

Fig. 4 shows the fraction of blue galaxies, as a function of cluster-centric
distance, from Andreon et al. (2005). These authors studied a volume-complete
sample of clusters (x-ray selected, to avoid the cluster-selection bias
mentioned in sec 2.1), at $z\sim0.35$, and compared it to several local cluster
samples. Cluster sizes have been measured with a good precision. An evolving
limiting magnitude  is used to select galaxies to be compared. Similarly, an
evolving $\Delta (B-V)$ is used to divide galaxies in colour classes. The
statistical analysis relies on the rigorous computation introduced in
d'Agostini (2004). Figure 4 shows several interesting results. First of all,
the cluster affects the properties of the galaxies up to two virial radii at
$z\sim0.35$. Second, the equality of the $z\sim0.35$ and $z\sim0.0$ fractions
of blue galaxies implies that during the last 3 Gyrs no evolution of the
fraction of blue galaxies, from the cluster core to the field value, is seen.
Since the authors adopt a definition of blue that accounts for the reduced age
of the universe at high redshift,  the evolution of which the authors talk
about is  beyond star aging.  Finally, the agreement of the radial profiles of
the fraction of blue galaxies at $z=0$ and $z\sim0.35$ implies that the pattern
infall did not change over the last 3 Gyr, or, at least, its variation has no
observational  effect on the studied quantity. 

At this meeting, Loh et al. (2005) presented an identical plot, but 
for a composite cluster formed by re-scaling and staking a large
cluster sample, without found the radial gradient shown in Fig.
4. As the authors noted, their lack of detection could be due to their
noisy measure of the cluster size, needed to rescale clusters during the
staking, because a noise measurement smoothes out any possible gradient.
The uncertainty on the cluster size, needed to measure the fraction of blue galaxies
at a reference radius (sec 2.2) also affects their claim about the existence of a
Butcher-Oemler effect.

To summarize, its seems that time is ripe for a reliable determination
of the colour and luminosity evolution of galaxies and how they
depend on environment, as also testified by several results
presented at this meeting.

\acknowledgements{It is a pleasure to thank all the colleagues that
helped me all along these years on the topics discussed in this paper. I
thank co-authors of my latest paper on the BO effect for
allowing me to present Fig. 4 in advance of publication.}

\vfill 
\end{document}